%% file: main.tex
\documentclass[conference,compsoc]{IEEEtran}
\IEEEoverridecommandlockouts

\usepackage{cite}
\usepackage{amsmath,amssymb,amsfonts}
\usepackage{graphicx}
\usepackage{textcomp}
\usepackage{xcolor}
\usepackage{hyperref}
\usepackage{url}
\usepackage{xurl}
\usepackage{ifthen}
\usepackage[T1]{fontenc}
\usepackage{minted}
\usepackage{tikz}
\usetikzlibrary{tikzmark,calc,shapes.geometric,arrows.meta,positioning,fit,backgrounds}
\usepackage{pifont}
\usepackage{pdfpages}
\usepackage{booktabs}
\usepackage{colortbl}
\usepackage{multirow}
\usepackage{hhline}
\usepackage{adjustbox}
\usepackage{float}
\usepackage{enumitem}
\usepackage{threeparttable}

\usetikzlibrary{positioning,arrows.meta}

\usepackage{listings}
\makeatletter\DeclareRobustCommand{\_}{\ifmmode\nfss@text{\textunderscore}\else\textunderscore\allowbreak\fi}\makeatother
\begin{document}

\title{Finding Memory Leaks in C/C++ Programs via Neuro-Symbolic Augmented \\ Static Analysis
}

\author{
\IEEEauthorblockN{%
Huihui Huang\textsuperscript{1},
Jieke Shi\textsuperscript{1}\textsuperscript{*}\thanks{\textsuperscript{*}Corresponding author.},
Bo Wang\textsuperscript{2},
Zhou Yang\textsuperscript{3},
David Lo\textsuperscript{1}}
\IEEEauthorblockA{%
\textsuperscript{1}Singapore Management University, Singapore \\
\textsuperscript{2}Beijing Jiaotong University, Beijing, China \\
\textsuperscript{3}University of Alberta \& Alberta Machine Intelligence Institute, Alberta, Canada \\
\{hhhuang, jiekeshi, davidlo\}@smu.edu.sg,\quad
wangbo\_cs@bjtu.edu.cn,\quad
zy25@ualberta.ca}
}

\newcommand{\circled}[1]{%
  \tikz[baseline=(char.base)]{
    \node[shape=circle,fill=black,inner sep=1pt](char)
      {\color{white}\tiny #1};
  }%
}
\newcommand{\hlcode}[1]{\colorbox{yellow!30}{\strut #1}}
\newcommand{\thickhline}{\Xhline{2\arrayrulewidth}}
\newcommand{\toolname}{\textsc{MemHint}}

\newboolean{showcomments}
\setboolean{showcomments}{true}
\ifthenelse{\boolean{showcomments}}
 { \newcommand{\mynote}[2]{
      \fbox{\bfseries\sffamily\scriptsize#1}
        {\small$\blacktriangleright$\textsf{\emph{#2}}$\blacktriangleleft$}}}
        { \newcommand{\mynote}[2]{}}
\definecolor{ForestGreen}{RGB}{0,180,0}
\newcommand{\todoforestgreen}[1]{\textcolor{ForestGreen}{\textbf{#1}}}
\newcommand{\todoblue}[1]{\textcolor{blue}{\textbf{#1}}}
\newcommand{\todored}[1]{\textcolor{red}{\textbf{#1}}}
\newcommand{\hh}[1]{\mynote{Huihui}{\todored{#1}}}
\newcommand{\rw}[1]{\textcolor{black}{#1}}

\maketitle

\begin{abstract}
Memory leaks remain prevalent in real-world C/C++ software. Static analyzers such as CodeQL provide scalable program analysis but frequently miss such bugs because they cannot recognize project-specific custom memory-management functions and lack path-sensitive control-flow modeling. We present \toolname{}, a neuro-symbolic pipeline that addresses both limitations by combining LLMs' semantic understanding of code with Z3-based symbolic reasoning. \toolname{} parses the target codebase and applies an LLM to classify each function as a memory allocator, deallocator, or neither, producing function summaries that record which argument or return value carries memory ownership, extending the analyzer's built-in knowledge beyond standard primitives such as \texttt{\small malloc} and \texttt{\small free}. A Z3-based validation step checks each summary against the function's control-flow graph, discarding those whose claimed memory operation is unreachable on any feasible path. The validated summaries are injected into CodeQL and Infer via their respective extension mechanisms. Z3 path feasibility filtering then eliminates warnings on infeasible paths, and a final LLM-based validation step confirms whether each remaining warning is a genuine bug. On eight real-world C/C++ projects totaling over 3.6M lines of code, \toolname{} detects 54 unique memory leaks, all confirmed or fixed, at approximately \$1.76 per detected bug, compared to 19 by vanilla CodeQL and 3 by vanilla Infer.
\end{abstract}




\section{Introduction}

C/C++, as programming languages with explicit memory management, require developers to manually deallocate memory regions that are no longer needed. Failure to do so results in \textit{memory leaks}: unreleased memory cannot be reclaimed by the operating system, gradually depleting system resources and potentially causing system failures~\cite{10.1145/3456727.3463783,299890} or exploitable security vulnerabilities~\cite{CWE401,8812075}. Recent high-profile incidents underscore this risk: a memory leak in Windows 11's update service caused unbounded RAM consumption of up to 20GB, rendering affected systems unresponsive~\cite{yahoo}, while a heap memory leak in Ollama enabled unauthenticated attackers to extract API keys and user conversations, potentially affecting $\sim$300,000 servers worldwide~\cite{cyeraBleedingLlama}. With over 1,800 CVEs (Common Vulnerabilities and Exposures) attributed to memory leaks as of May 2026~\cite{opencve}, effective automated detection of such defects remains a pressing need for software reliability and security.

Memory leaks are notoriously difficult to detect. Dynamic analysis approaches such as fuzzing~\cite{MemLock,10.1145/3551349.3561161,han2018enhancing,11391877} can identify leaks along executed paths but impose substantial runtime overhead that limits scalability for large-scale or long-running software~\cite{10.1145/3133956.3134046,AddressSanitizer}, and may fail to exercise the specific paths needed to trigger a leak within a practical time budget (as discussed in Section~\ref{sec:motivation}). Static analysis is a more attractive alternative, as it can reason over all program paths without requiring execution. Static analyzers such as CodeQL~\cite{codeql_github} and Infer~\cite{fb_infer_github} perform interprocedural dataflow analysis to trace how memory is allocated and deallocated, detecting mismatched or missing deallocations, and have seen wide industrial adoption~\cite{10.1145/3338112,githubGitHubActions}. However, these analyzers {\it lack semantic understanding of custom memory management and path feasibility}, resulting in both missed detections and spurious warnings.

In real-world C/C++ projects, developers often implement custom variants of Memory Management (MM) functions rather than directly calling standard primitives such as \texttt{\small malloc} and \texttt{\small free}. This practice is especially prevalent in large open-source codebases that span many modules from diverse contributors~\cite{299890,lyu2022goshawk,shemetova2025lamed,liu2024}, where a single project may define hundreds of custom memory allocation and deallocation functions (e.g., \texttt{\small tree\_init} and \texttt{\small policy\_tree\_free} in OpenSSL~\cite{openssl}). Effective leak detection thus requires identifying these custom functions and their allocation/deallocation pairings.

Unfortunately, existing static analyzers such as CodeQL by default operate with a fixed list of standard primitives, remaining blind to custom MM functions and silently missing leaks mediated by them. Prior work~\cite{242050,6671326,299890,lyu2022goshawk,liang2025leakguard} resorts to heuristics or natural language processing (NLP) techniques that infer memory management roles from function names or documentation, yet these heuristics do not check whether the inferred allocation or deallocation is reachable on any feasible intraprocedural path. A function may contain a syntactic call to \texttt{\small free} inside a branch that is never satisfiable; labeling such a function as a deallocator will cause the analyzer to wrongly assume the memory can be freed through it, suppressing genuine leak reports. Even when custom MM functions are correctly identified, the analyzers themselves remain path-insensitive: they often over-approximate program behavior and report warnings along paths that are infeasible under actual runtime conditions, flooding developers with false alarms that demand costly manual triage~\cite{10.1145/3715729,10.1145/3494521,10.1145/3510003.3510214}.

Our key insight is that addressing the limitations above requires both semantic understanding and symbolic reasoning: identifying custom MM functions calls for comprehending developer intent from code context, while validating those identifications and filtering false positives demands ownership-aware symbolic analysis of program control flow. We realize this insight in \toolname{}, which is a neuro-symbolic pipeline. For the neuro component, we leverage Large Language Models (LLMs), which have shown strong capabilities across various programming tasks~\cite{10.1145/3695988,10.1145/3747588,10.1145/3708525} and cybersecurity tasks~\cite{10.1145/3769082,10.1145/3769676}, to infer the memory management roles of functions. However, LLMs are susceptible to hallucinations~\cite{10.1145/3728894,11121691} and may produce incorrect annotations. We thus complement them with a symbolic component that validates each LLM-generated annotation and filters analyzer output against path feasibility constraints using the Z3 Satisfiability Modulo Theories (SMT) solver~\cite{z3}. By composing the two, existing static analyzers can be augmented for more precise memory leak detection.

Concretely, \toolname{} operates in three stages. It first parses the target codebase to extract macros and functions along with their signatures, bodies, and callees. An LLM (Gemini 3 in our study) then analyzes each function and its callees to label it as a memory allocator, deallocator, or neither, producing a function summary that records the function's MM role and which argument or return value carries memory ownership. Since LLMs may produce incorrect annotations, \toolname{} further constructs a control-flow graph for each annotated function and encodes its path conditions into Z3 to check whether the claimed allocation or deallocation is reachable on any feasible path, discarding unsound annotations. In the second stage, the validated MM function summaries are injected into static analyzers (CodeQL and Infer) via their respective extension mechanisms. In the final stage, \toolname{} uses Z3 to check whether each reported leak path is feasible and discards warnings on infeasible paths; a final LLM-based validation step confirms whether each remaining warning is a genuine bug.

We evaluate \toolname{} on eight widely-used open-source C/C++ projects totaling over 3.6 million lines of code. \toolname{} detects 54 unique memory leak bugs, all of which have been confirmed or fixed by project maintainers or through our submitted patches. In comparison, current open-source and industrial-strength static analyzers detect far fewer: the best-performing baseline, vanilla CodeQL, finds only 19, while vanilla Infer detects only 3, Semgrep~\cite{semgrep} only 2, and the recent state-of-the-art academic tool LeakGuard~\cite{liang2025leakguard} finds 20. We construct a human-labeled dataset of 1{,}532 functions; evaluation on it shows that \toolname{} achieves 100\% precision and recall on reachable memory management function identification, substantially outperforming LeakGuard (34.0\% precision, 52.2\% recall). An ablation study confirms that each pipeline component independently contributes to the overall effectiveness. Additionally, the total cost per detected bug is approximately \$1.76, and the static analysis portion of the pipeline runs 4.9--5.7$\times$ faster than LeakGuard. These results demonstrate that \toolname{} effectively bridges the gap between semantic understanding and symbolic reasoning, enabling precise and scalable memory leak detection in real-world C/C++ software.

Our paper makes the following contributions: 
\begin{itemize}[leftmargin=*] 

\item We frame memory leak detection as a neuro-symbolic task: LLMs infer custom memory-management functions from code context, while Z3 validates their reachability and filters infeasible leak warnings.
    
\item We show that LLM-generated MM function summaries can faithfully encode allocator/deallocator semantics and, once injected into static analyzers, substantially boost their detection performance.

\item We implement \toolname{}, a neuro-symbolic pipeline that generates and validates function summaries via LLMs and Z3, injects them into CodeQL and Infer, and leverages Z3 and a final LLM pass to suppress false alarms. Our replication package is available at \url{https://github.com/jiekeshi/MemHint}.

\item We evaluate \toolname{} on eight real-world C/C++ projects and detect 54 unique memory leak bugs (all confirmed or fixed), outperforming vanilla CodeQL (19), vanilla Infer (3), Semgrep (2), and LeakGuard (20).

\end{itemize}

\section{Motivating Example}
\label{sec:motivation}


\begin{figure}[t!]
\centering
    \includegraphics[page=1,width=\linewidth]{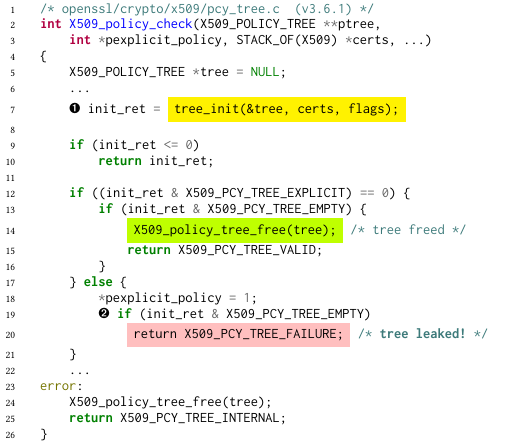}

\caption{A memory leak in OpenSSL (v3.6.1)~\cite{openssl}.
\texttt{\small tree\_init} allocates a policy tree \ding{202}.
The green path frees it correctly, but at \ding{203} the function returns without freeing \texttt{tree} (red).
24 hours of fuzzing with LeakSanitizer failed to trigger this path.}

\label{fig:motivating-example-openssl}
\end{figure}

This section motivates \toolname{} by showing why current approaches fall short on memory leak detection.

Prior work has explored dynamic analysis approaches~\cite{8539206}, particularly fuzzing~\cite{MemLock,10.1145/3551349.3561161,han2018enhancing,11391877}, for detecting memory leaks. These techniques combine coverage-guided input generation with runtime sanitizers (e.g., LeakSanitizer~\cite{LeakSanitizer}) that monitor memory allocation/deallocation events along executed paths. However, dynamic approaches can only detect leaks along paths that the fuzzer actually exercises, and may silently miss leaks that require specific branch combinations that are difficult to reach through crafted inputs. Figure~\ref{fig:motivating-example-openssl} illustrates this limitation with a confirmed memory leak in OpenSSL (v3.6.1), one of the most widely used cryptographic libraries~\cite{openssl}. The function \texttt{\small X509\_policy\_check()} allocates a policy tree via \texttt{\small tree\_init} (\ding{202}). When the tree is non-explicit and empty, the function correctly frees it before returning; yet when it is both \texttt{\small EXPLICIT} and \texttt{\small EMPTY} (\ding{203}), the function returns without freeing the tree, leaking the allocated memory. Triggering this path requires an input satisfying both conditions simultaneously, which a fuzzer may not systematically produce.

OpenSSL has been continuously fuzzed by Google's OSS-Fuzz~\cite{203944} since 2016, yet this bug, along with all other bugs found by \toolname{}, does not appear in its bug tracker. We further rebuilt OpenSSL with LibFuzzer~\cite{LibFuzzer} and LeakSanitizer~\cite{LeakSanitizer} enabled and ran the fuzzer for 24 hours; it executed over 9.8 billion test cases yet failed to detect this bug, suggesting that such path-dependent leaks are difficult to reliably detect via dynamic analysis. Static analysis is therefore a natural complement, as it can reason over all program paths without requiring execution. However, current static analyzers still suffer from several limitations that \toolname{} is designed to address.


\begin{figure}[t!]
\centering
   \includegraphics[page=1,width=\linewidth]{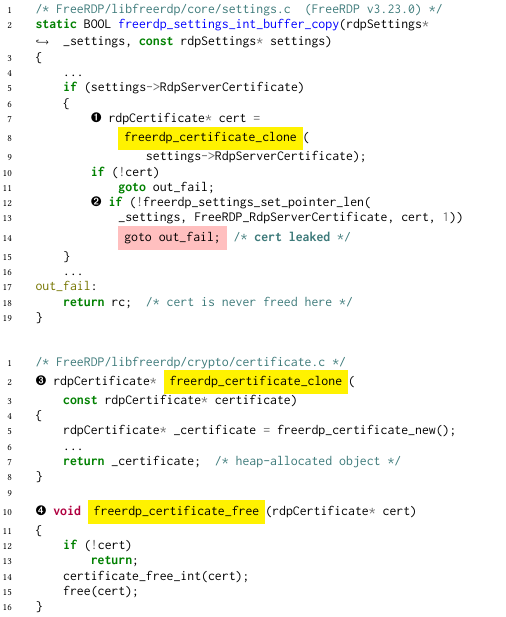}

\caption{A memory leak in FreeRDP (v3.23.0)~\cite{freerdp}. If \texttt{\small freerdp\_settings\_set\_pointer\_len} fails (\ding{203}), the object allocated by the custom allocator \texttt{\small freerdp\_certificate\_clone} (\ding{202}) is never freed. Static analyzers miss this bug as neither the allocator nor its paired deallocator (\ding{205}) is a known MM function.}

\label{fig:motivating-example-freerdp}
\end{figure}

The first and foremost limitation is that these tools cannot recognize custom memory management functions. Real-world software often implements project-specific MM functions that wrap standard primitives such as \texttt{\small malloc} and \texttt{\small free} behind multiple layers of abstraction~\cite{299890,shemetova2025lamed}. Although the underlying primitives are transitively reachable, static analyzers such as CodeQL~\cite{codeql_github}, Clang Static Analyzer~\cite{clang_static_analyzer}, and Infer~\cite{fb_infer_github} do not fully inline every function but instead rely on predefined function summaries, which are abstract descriptions of how a function affects memory (e.g., ``\texttt{\small malloc} takes a size argument and returns a heap-allocated pointer''), for a fixed set of standard primitives. For custom wrappers outside this set, the analyzer has no summary and thus cannot determine whether a call allocates or frees memory. Even if the analyzer attempted to look inside these wrappers for standard primitives, multi-layer delegation and conditional logic would break dataflow tracking across call boundaries. As a result, the analyzer cannot establish that a wrapper's return value carries heap ownership or that calling a wrapper constitutes a deallocation, leaving allocation/deallocation pairings invisible and memory leaks undetected.

Figure~\ref{fig:motivating-example-freerdp} illustrates this limitation with a real-world memory leak in FreeRDP (v3.23.0), a widely used open-source Remote Desktop Protocol implementation~\cite{freerdp}. \texttt{\small freerdp\_certificate\_clone} (\ding{202}, \ding{204}) is a custom allocator that deep-copies a certificate into heap memory. When \texttt{\small freerdp\_settings\_set\_pointer\_len} fails (\ding{203}), execution jumps to \texttt{\small out\_fail} without calling the paired deallocator \texttt{\small freerdp\_certificate\_free} (\ding{205}), leaking the certificate. Existing static analyzers miss this bug entirely: neither of the two MM functions, \texttt{\small freerdp\_certificate\_clone} and \texttt{\small freerdp\_certificate\_free}, has a predefined summary in their memory model, and the multi-layer delegation within these wrappers prevents the analyzer from tracking the underlying primitives (e.g., \texttt{\small freerdp\_certificate\_free} $\rightarrow$ \texttt{\small certificate\_free\_int} $\rightarrow$ \texttt{\small X509\_free} $\rightarrow$ \texttt{\small free} in one of FreeRDP's deallocators) across call boundaries.

Several studies have attempted to identify custom MM functions using heuristics, machine learning, or natural language processing (NLP) over function names, documentation, and usage patterns~\cite{242050,299890,liang2025leakguard,lyu2022goshawk,liu2014pf,liu2015pairminer}. However, these methods rely on surface-level signals such as function names, which are unreliable in practice: a function named \texttt{\small cleanup} may or may not free memory depending on its implementation, and the same name can carry different semantics across projects. Crucially, these methods do not validate whether the allocation or deallocation operation is actually \emph{reachable} under the function's control flow. A function may contain a syntactic call to \texttt{\small free} inside a branch that is never satisfiable, yet a heuristic or NLP-based approach would still annotate it as a deallocator, introducing false annotations that actively mislead the analyzer. Symbolic validation of MM function annotations is therefore necessary but absent from prior work.

\begin{figure*}[t!]
    \centering
    \includegraphics[page=1,width=\linewidth]{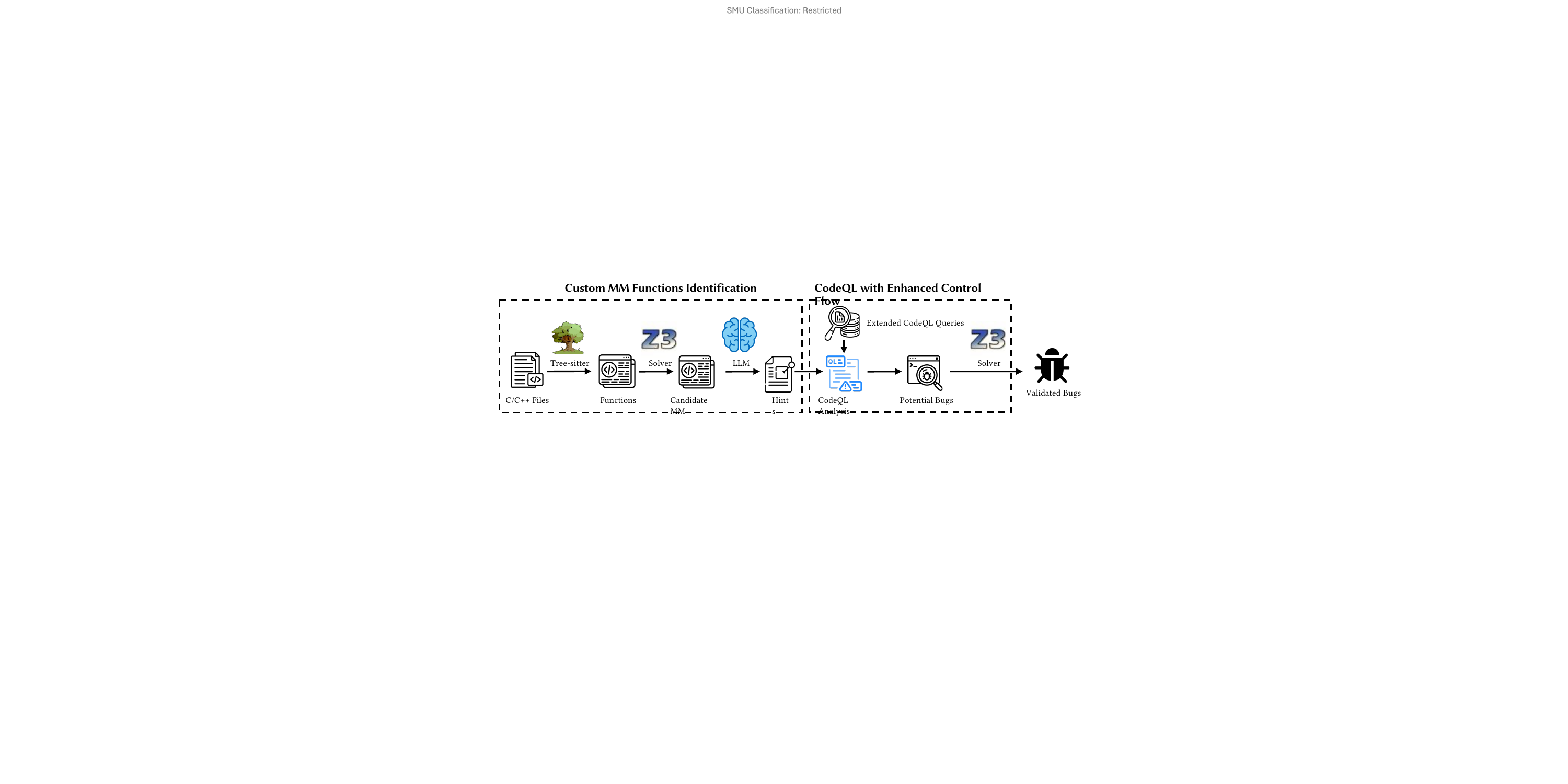}
\caption{
Overview of the \toolname{} pipeline.
\textbf{Stage~1 (Summary Generation)}: extract metadata, generate summaries with an LLM, and validate with Z3 SMT solver.
\textbf{Stage~2 (Summary-Augmented Analysis)}: inject validated summaries into static analyzers (e.g., CodeQL, Infer).
\textbf{Stage~3 (Warning Validation)}: filter infeasible warnings with Z3 and validate remaining ones using an LLM.
}
    \label{fig:workflow}
\end{figure*}

The final limitation lies in static analyzers themselves. Even when a static analyzer is given an accurate memory model with comprehensive MM functions and their summaries, it may still report spurious warnings~\cite{10.1145/3715729,10.1145/3494521,10.1145/3510003.3510214,guo2023mitigating}. These tools are largely path-insensitive: they merge control-flow branches into a single abstract flow and cannot determine whether a reported warning is reachable under actual program semantics. A memory leak that appears along one branch may in fact be guarded by conditions that make the path infeasible at runtime, yet the tool reports it regardless. For example, two branches may contain mutually exclusive conditions that the analyzer fails to recognize, causing it to report a memory leak on a path that can never execute. Without a mechanism to filter out such warnings with infeasible paths, users must manually triage a large number of false alarms, significantly reducing the practical utility of the tool~\cite{li2023comparison,muske2016survey,guo2023mitigating}.

\toolname{} addresses these limitations by combining LLMs' semantic understanding of code with symbolic reasoning through constraint solving. \toolname{} first uses an LLM to infer the memory management roles of custom functions from their source code and callees, producing function summaries that extend the analyzer's built-in knowledge beyond the standard primitives. Each summary is then validated by Z3 against the function's intraprocedural control-flow graph, ensuring that the identified allocation or deallocation is reachable on a feasible path. In addition, \toolname{} encodes each analyzer warning as a path constraint and uses Z3 to discard those on infeasible paths, reducing false alarms before they reach the developer. Together, these steps form a neuro-symbolic pipeline in which semantic understanding and formal reasoning reinforce each other at every stage.

\section{\toolname{}}

As shown in Figure~\ref{fig:workflow}, \toolname{} operates in three stages. In \emph{Stage~1 (Summary Generation)}, \toolname{} extracts each function's metadata (e.g., signature, body, callees), filters functions via pointer-type heuristics, uses an LLM to produce structured MM function summaries, and validates each summary with the Z3 SMT solver~\cite{z3}. In \emph{Stage~2 (Summary-Augmented Analysis)}, the validated summaries are injected into CodeQL and Infer via their extension mechanisms, while in \emph{Stage~3 (Warning Validation)}, \toolname{} applies Z3-based path feasibility checking and uses an LLM to filter false positives.

\subsection{Stage~1: Summary Generation}
\label{subsec:stage1}

\noindent\textbf{Phase~1: Code Extraction.}
\toolname{} begins by parsing the target codebase to extract metadata for each function. Specifically, we use Tree-sitter~\cite{tree_sitter}, a widely-used parsing library, to extract each function's name, return type, parameter list with types, full source body, and the set of direct callee names appearing in call expressions. We also extract C preprocessor macros that expand to function-like constructs, converting them into the same record format so they can be analyzed for memory-management semantics. In addition, we collect pointer-typedef aliases, which are used in the subsequent pre-filtering step to recognize pointer types that do not contain an explicit \texttt{\small *} in the function signature.

\noindent\textbf{Phase~2: LLM Summary Generation.}
In this phase, \toolname{} classifies each candidate function as an allocator, a deallocator, or neither, and produces a structured \emph{function summary} for each identified MM function. A function summary is an analyzer-agnostic annotation that records the function's MM role (allocator or deallocator) and which argument or return value carries memory ownership; it can be directly translated into the extension formats of different static analyzers.
To reduce LLM cost, we apply a lightweight pre-filter and retain only functions whose signatures involve pointer types (e.g., return type or parameters containing \texttt{\small *} or pointer-typedef aliases). Macros are always retained, as they may expand to pointer operations not visible in their signature. Entry-point and test functions (e.g., \texttt{\small main}, \texttt{\small wmain}, and functions whose names contain \texttt{\small test}) are excluded.

Each remaining function is analyzed by an LLM using its signature, full source body, and the source code of up to five direct callees as context (the full prompt template is given in Appendix~\ref{appendix:prompt-classify}). Providing callee context is important because many project-specific memory-management functions delegate to helper functions, making their semantics non-local. Functions classified as neither are discarded. For each identified MM function, the LLM produces a structured \emph{function summary} containing three fields: (1)~\texttt{\small name}, the function's name identifier; (2)~\texttt{\small role}, the function's memory management role, either \textsc{Allocator} or \textsc{Deallocator}; and (3)~\texttt{\small target}, which specifies \texttt{\small return} for allocators (indicating the return value carries heap ownership) or \texttt{\small arg}$i$ for deallocators (indicating the $i$-th argument is freed). For example, given \texttt{\small freerdp\_certificate\_clone} in Figure~\ref{fig:motivating-example-freerdp}, the LLM produces \texttt{\small \{name: freerdp\_certificate\_clone, role: Allocator, target: return\}}, as it calls \texttt{\small freerdp\_certificate\_new} and returns the newly allocated pointer. Conversely, \texttt{\small freerdp\_certificate\_free} yields \texttt{\small \{role: Deallocator, target: arg0\}}, as it frees its first argument.

\input{z3_validation_figure}

\noindent\textbf{Phase~3: Z3-Based Summary Validation.}
LLM-generated summaries require validation before use for two reasons. First, the LLM may misclassify functions; e.g., allocator-like names may not imply actual allocation, or control flow may obscure deallocation reachability. Second, operations such as \texttt{\small malloc} or \texttt{\small free} may occur only on infeasible paths, rendering the summary invalid. \toolname{} validates each summary using the Z3 SMT solver~\cite{z3}. Specifically, \toolname{} constructs an intraprocedural control-flow graph (CFG), encodes path conditions as constraints, and checks whether a feasible path exists on which the claimed allocation or deallocation occurs in the corresponding MM function. Summaries without such a feasible path are discarded, and the remaining ones are retained as \emph{validated summaries}. This step also leverages callee information from Phase~2 to handle interprocedural cases by following transitive call chains up to a bounded depth, correctly validating summaries involving transitive delegation (e.g., a wrapper that calls another wrapper before reaching \texttt{\small malloc}), parameter aliasing (e.g., assigning a parameter to a local variable before freeing it), and indirect parameter derivation (e.g., freeing the object that a pointer-to-pointer parameter refers to). Note that the alias analysis is intraprocedural and identifier-based: it tracks assignments of the form \texttt{\small v = p} and \texttt{\small v = *p} together with their transitive closure, which suffices to recognize a \texttt{\small free(v)} as freeing parameter \texttt{\small i} when \texttt{\small v} is derived from it. Transitive allocator and deallocator validation follows the project call graph up to a fixed depth bound (10 in our implementation), which ensures termination on recursive and mutually recursive code. All candidate summaries are generated before validation, and each callee is validated recursively down to primitives or reuses a cached result, so the order does not matter.

We first describe how the CFG is constructed. For each function to be validated, \toolname{} parses its source and builds an intraprocedural control-flow graph (CFG) with a designated \textsc{Entry} node and one or more \textsc{Exit} nodes, as illustrated in Figure~\ref{fig:z3-validation}. Each node is assigned a semantic type: \textsc{Alloc} for calls to allocation primitives (e.g., \texttt{\small malloc}, \texttt{\small calloc}, \texttt{\small realloc}, \texttt{\small strdup}), \textsc{Free} for calls to deallocation primitives (e.g., \texttt{\small free}, \texttt{\small delete}), \textsc{Return} for function exits, \textsc{Branch} for conditional control flow (e.g., \texttt{\small if}/\texttt{\small else}, \texttt{\small switch}), \textsc{Assign} for assignments, and \textsc{Deref} for pointer dereferences. Sequential statements are connected by single-successor edges, while each \textsc{Branch} node has two outgoing edges corresponding to the \emph{true} and \emph{false} outcomes, guarded by a Boolean variable $b_v$ and its negation $\neg b_v$.

For a function labeled \textsc{Allocator}, \toolname{} checks whether there exists a feasible CFG path on which (i)~a known allocation primitive (or a transitively identified allocator wrapper) is invoked, and (ii)~the allocated value reaches the function's return site without being overwritten, released, or stored into a global table or a container. As illustrated in Figure~\ref{fig:z3-validation} (a)--(b), a summary is retained only if such a feasible path exists.

Let $G = (V, E)$ denote the intraprocedural CFG, where each branch node $v$ introduces a Boolean variable $b_v$ for its condition; branch nodes that test the same condition, with no write to its operands in between, share one variable, and a negated test uses its negation. A path $\pi = (n_0, n_1, \ldots, n_k)$ from \textsc{Entry} to a \textsc{Return} node is \emph{feasible} if its path condition $\varphi(\pi) = \bigwedge_{v \in \mathit{branches}(\pi)} l_v$ is satisfiable, where $l_v = b_v$ if $\pi$ takes the true branch at $v$ and $l_v = \neg b_v$ otherwise. Let $A(\pi)$ denote the set of \textsc{Alloc} nodes on $\pi$, and let $\mathit{reaches}(a, r, \pi)$ be true when the value allocated at node $a$ reaches the return site $r$ along $\pi$ without being overwritten, freed, or stored into a global table or a container. An allocator summary is \emph{valid} if and only if
\begin{equation}
\label{eq:alloc-valid}
\begin{aligned}
\exists\, \pi \in \mathit{Paths}(G),\; \exists\, a \in A(\pi) : & \\
\textsc{Sat}\bigl(\varphi(\pi) \wedge \mathit{reaches}(a, r, \pi)\bigr). &
\end{aligned}
\end{equation}
If the constraint is satisfiable (SAT) for at least one path, the summary is retained; if it is unsatisfiable (UNSAT) for all candidate paths, the summary is discarded.

For a \textsc{Deallocator} summary associated with argument index~$i$, \toolname{} checks whether there exists a feasible CFG path on which the value of parameter~$i$, or a value derived from it, is passed to a known deallocation primitive. As illustrated in Figure~\ref{fig:z3-validation} (c)--(d), a summary is retained only if such a path frees the target value. To track such flows, \toolname{} uses a lightweight alias analysis that records variables derived from parameter~$i$ and propagates this information along \textsc{Assign} nodes. A \textsc{Free} node is considered a valid deallocation if its argument aliases parameter~$i$ under this analysis. To handle delegation patterns, \toolname{} additionally follows transitive call chains using a call graph constructed from the codebase. If a callee has already been identified as a deallocator or is itself undergoing validation, the analysis recursively follows the chain up to a bounded depth.

Let $F(\pi)$ denote the set of \textsc{Free} nodes occurring on path $\pi$, and let $\mathit{aliases}(f, i, \pi)$ be true when the argument freed at node $f$ is an alias of parameter~$i$ (or a value transitively derived from it) along $\pi$. A deallocator summary is \emph{valid} if and only if
\begin{equation}
\label{eq:dealloc-valid}
\begin{aligned}
\exists\, \pi \in \mathit{Paths}(G),\; \exists\, f \in F(\pi) : & \\
\textsc{Sat}\bigl(\varphi(\pi) \wedge \mathit{aliases}(f, i, \pi)\bigr). &
\end{aligned}
\end{equation}
If the constraint is SAT, the summary is retained and carried into Stage~2; otherwise (UNSAT), it is discarded.

\subsection{Stage~2: Summary-Augmented Analysis}
\label{subsec:stage2}

This stage injects the validated summaries into CodeQL~\cite{codeql_github} and Infer~\cite{fb_infer_github} and runs both analyzers to detect memory leaks. The validated summaries are \emph{analyzer-agnostic}: each summary specifies whether a function allocates or frees memory and which argument or return value carries ownership, without relying on any tool-specific representation (Appendix~\ref{appendix:summary-example} shows a concrete summary and its injected forms for CodeQL and Infer). We select CodeQL and Infer due to their complementary analysis foundations (declarative interprocedural dataflow and separation logic) and because both expose a programmable memory model; analyzers without such an interface, such as the Clang Static Analyzer, would require a separate integration.

For CodeQL, a declarative dataflow analysis engine that traces value propagation across function boundaries, \toolname{} injects the validated summaries through its extension mechanism, populating \texttt{\small allocation\allowbreak{}Function\allowbreak{}Model} and \texttt{\small deallocation\allowbreak{}Function\allowbreak{}Model}, which are CodeQL's built-in extension points for registering custom MM functions. However, CodeQL's default memory leak queries (e.g., \texttt{\small Memory\allowbreak{}Never\allowbreak{}Freed}) are largely path-insensitive: they merge control-flow branches into a single abstract flow and check only whether a deallocation exists somewhere after an allocation, without distinguishing which branch performs it. As a result, a leak that occurs on a specific branch, such as an error-handling path that returns early without freeing, may be missed if another branch does free the memory. To address this, \toolname{} augments CodeQL with extended queries that (i)~track deallocation on a per-branch basis, reporting a leak when any individual branch fails to free the allocated memory; (ii)~specifically model error-handling paths (e.g., \texttt{\small goto} cleanup labels and early returns after failed checks) as potential leak sites; and (iii)~detect cases where deallocation is guarded by a condition that may not always hold, leaving the memory unfreed on the alternative path.

For Infer, \toolname{} translates each validated summary into the configuration format of Infer's Pulse engine~\cite{fb_infer_github}, a memory and value analysis engine developed by Meta that is based on separation logic and tracks heap ownership along execution paths. Allocator summaries are registered via \texttt{\small -{}-pulse-model-alloc-pattern}, so that Infer treats their return values as newly allocated heap objects with caller ownership; deallocator summaries are similarly registered so that calls to these functions are recognized as deallocation operations.

\subsection{Stage~3: Warning Validation}
\label{subsec:stage3}

\input{z3_feasibility_figure}

\noindent\textbf{Phase~5: Z3-Based Path Feasibility Filtering.}
Static analyzers may report memory leaks along infeasible paths, i.e., paths that exist syntactically but are unreachable at runtime due to implicit invariants or mutually exclusive branch conditions. To address this, \toolname{} performs an intraprocedural analysis for each flagged function and uses Z3 to check whether a feasible execution path exists that exhibits the reported leak. Warnings without such a feasible path are discarded. 

Concretely, for each warning, \toolname{} parses the flagged function and constructs a CFG using the same node classification as in Phase~3. Allocation and deallocation operations include both standard primitives (e.g., \texttt{\small malloc}, \texttt{\small free}) and custom allocators/deallocators recorded in validated summaries (e.g., \texttt{\small freerdp\_certificate\_clone} in Figure~\ref{fig:motivating-example-freerdp}). \toolname{} then encodes the CFG as a set of constraints over three Boolean state variables that track the status of pointer $p$ along each path:
\begin{itemize}[nosep,leftmargin=*]
\item $\mathit{alloc}(n)$: memory for $p$ has been allocated at or before node $n$;
\item $\mathit{freed}(n)$: memory for $p$ has been freed at/before node $n$;
\item $\mathit{escaped}(n)$: ownership of $p$ has been transferred at or before node $n$ (e.g., via return, global store, or ownership-sink call).
\end{itemize}

Only allocation, deallocation, and escape-related nodes update these states; all other nodes propagate them unchanged, keeping the encoding lightweight. The encoding does not model pointer values, heap cells, or array theories. Because Phase~3 has already labeled custom allocator and deallocator wrappers as \textsc{Alloc} or \textsc{Free}, Phase~5 can stay intraprocedural: when the function under analysis calls a deallocator wrapper, \toolname{} treats the call site itself as a \textsc{Free} node, without analyzing the wrapper's body. For each allocation site, \toolname{} checks whether there exists a feasible path from the allocation to a function exit along which the pointer is neither freed nor ownership-transferred. As shown in Figure~\ref{fig:z3-feasibility}, such a path reaches a \textsc{Return} node without encountering a matching deallocation.

To enforce a single concrete path, \toolname{} introduces a Boolean edge variable $e_{(u,v)}$ for each CFG edge and constrains each node to receive at most one incoming edge
\begin{equation}
\label{eq:single-path}
\forall\, n \in V :\; \textsc{AtMost}_1\bigl(\{e_{(u,n)} \mid (u,n) \in E\}\bigr).
\end{equation}
Node reachability is defined as $\mathit{reach}(n) \Leftrightarrow \bigvee_{(u,n) \in E} e_{(u,n)}$, with $\mathit{reach}(\textsc{Entry}) = \top$.
The three state variables are propagated along the selected path. At each non-special node, the state is inherited from the chosen predecessor
\begin{equation}
\label{eq:state-propagate}
\mathit{state}(n) = \bigvee_{(u,n) \in E} \bigl(e_{(u,n)} \wedge \mathit{state}(u)\bigr).
\end{equation}
Since at most one incoming edge is active (Equation~\ref{eq:single-path}), the disjunction effectively selects the state from the unique chosen predecessor.
At \textsc{Alloc}, \textsc{Free}, and escape nodes, the corresponding state variable is additionally set to true when the node is reachable.
A memory leak is \emph{feasible} if and only if
\begin{equation}
\label{eq:leak-feasible}
\begin{aligned}
\exists r \in \mathit{Exits}(G): \\
\textsc{Sat}\bigl(\mathit{reach}(r) \land \mathit{alloc}(r) \land \neg \mathit{freed}(r) \land \neg \mathit{escaped}(r)\bigr).
\end{aligned}
\end{equation}
If Z3 returns \textsc{Sat}, the warning is retained; otherwise (\textsc{Unsat}), it is discarded as a false alarm.
All remaining warnings are forwarded to Phase~6. Note that our CFG models conditional, return, and goto control flow but is acyclic: each loop body (\texttt{\small for/while/do}) is modeled as executing at most once. Phase~5 tracks the pointer variable that receives the allocation; a warning whose allocation cannot be located in the CFG in this form cannot be encoded and is discarded, for example when the allocated pointer is passed straight into another call and never held by a variable. The leak patterns that dominate our setting, namely allocations on error branches and frees hidden behind custom wrappers (Figures~\ref{fig:motivating-example-openssl} and~\ref{fig:motivating-example-freerdp}), arise from control flow rather than loop iteration; loop-carried leaks are out of the current scope and discussed in Section~\ref{sec:limitation}.

\noindent\textbf{Phase~6: LLM-Based Warning Validation.}
After Z3-based feasibility filtering, the remaining warnings correspond to feasible execution paths but may still include residual false positives that intraprocedural symbolic analysis alone cannot eliminate.
\rw{In particular, the Z3 encoding captures path feasibility but not higher-level semantics such as callback-based deallocation or project-specific ownership conventions (e.g., borrowed pointers or non-owning references). For instance, in systems like Redis, certain APIs return pointers to internally managed data structures whose ownership remains with the library, so the caller is not responsible for deallocation; static analyzers nonetheless flag the missing free as a leak. Such conventions are encoded in API documentation rather than in code structure, placing them beyond the reach of constraint-based reasoning.
}

For each candidate warning, \toolname{} constructs a prompt that includes (i)~the full source code of the function, (ii)~the warning type (memory leak) together with the warning message and location, and (iii)~the analysis path reported by the static analysis tool (i.e., the CodeQL dataflow path or Infer error trace), with code snippets at each program point. The LLM is asked to determine whether the reported warning corresponds to a real bug or a false positive, together with a brief justification. Warnings classified as false positive are discarded, and the remaining ones are reported as the final output. The full prompt is given in Appendix~\ref{appendix:prompt-validate}.

\section{Evaluation}

We evaluate \toolname{} on eight real-world C/C++ projects, manually inspecting every warning and submitting confirmed cases to upstream maintainers. Our evaluation answers four questions: (i)~the \emph{Bug-Finding Effectiveness} of \toolname{}, measured by discovered bugs and comparison with prior tools; (ii)~the \emph{Summary Quality} of inferred MM-function annotations, measured by precision and recall; (iii)~the \emph{Component Contribution} of each pipeline stage to final detection performance; and (iv)~the \emph{Efficiency} of \toolname{}, measured by runtime and cost.

\subsection{Experiment Setup}

\rw{In Phase~2 (LLM summary generation), we use Gemini 3 Flash, while in Phase~6 (LLM warning validation), we use Gemini 3.1 Pro.} We choose a lightweight model for summary generation because it needs to process a large number of functions, where efficiency and cost are critical, whereas warning validation requires higher accuracy and thus benefits from a stronger model. Prompt templates and batching details are given in Appendix~\ref{appendix:prompts}.

We use CodeQL CLI version~2.25.1, Infer version~1.2.0, and Z3 version~4.15.4. We evaluate \toolname{} on eight widely used open-source C/C++ projects: Vim, tmux, OpenSSL, Redis, FreeRDP, curl, FFmpeg, and the
\texttt{linux/drivers/staging} subsystem of the Linux kernel, which we refer to
as \texttt{linux/staging} hereafter. We selected projects used by our baselines, prioritizing actively maintained and widely used projects across diverse domains. For each project, we run the full \toolname{} pipeline (Stages~1--3) on the entire codebase.

\subsection{Bug-Finding Effectiveness}
\input{tab_results}
Table~\ref{tab:results} summarizes the bug detection results of \toolname{} across all eight projects. In total, \toolname{} detects 54 unique memory-safety bugs, all of which have been confirmed or fixed by project maintainers or through our submitted patches. As CodeQL and Infer both receive the same validated summaries, they share 21 overlapping findings that both analyzers can detect once project-specific memory models are provided. However, CodeQL contributes 44 findings and Infer contributes 31, leaving 23 bugs found only by CodeQL and 10 found only by Infer. The non-overlap reflects differences in analysis strategy: CodeQL's declarative dataflow queries and Infer's bi-abductive reasoning capture complementary patterns. Summarizing a function does not hide leaks inside it, as its body is still analyzed: e.g., \toolname{} labels tmux's \texttt{\small file\_get\_path} as an allocator and still reports a leaked inner allocation in it.

\noindent \textbf{Comparison with Baselines.} We compare \toolname{} against four baselines: vanilla CodeQL~\cite{codeql_github}, vanilla Infer~\cite{fb_infer_github}, LeakGuard~\cite{liang2025leakguard}, and Semgrep~\cite{semgrep}. LeakGuard is a state-of-the-art memory-leak detection approach that infers paired memory-management functions and integrates them into static analysis. We use the authors' released implementation. Semgrep is a widely used source-level static analyzer that does not require compilation; we include it to provide a comparison with a pattern-based approach operating at a different abstraction level. We use Semgrep version~1.156.0 with the Pro engine enabled.

\input{tab_baseline}

Table~\ref{tab:baseline_comparison} reports the number of detected bugs per project for each tool. Because exhaustively validating all warnings reported by the baselines is infeasible, our comparison only checks whether each baseline can recover the bugs identified and manually validated in our study, rather than searching for additional baseline-only bugs. All bugs found by the baseline tools are strict subsets of those found by \toolname{}, and 20 bugs are uniquely found by \toolname{}. Vanilla CodeQL detects 19 bugs; notably, running CodeQL with our extended queries but \emph{without} summary injection produces the same 19 bugs (marked with $\ast$ in Table~\ref{tab:baseline_comparison}), indicating that the additional detections are attributable to the summaries rather than the enhanced queries alone. Vanilla Infer detects only 3 bugs. Without summaries, neither tool can track object lifetimes through project-specific allocators and deallocators, missing a substantial number of leaks. LeakGuard detects 20 bugs; although it identifies custom memory-management functions, it misses leaks that require deeper interprocedural reasoning or that arise from complex control-flow patterns. Semgrep detects only 2 bugs, as its purely syntactic analysis lacks the data-flow and path-sensitive reasoning required to detect interprocedural leaks involving custom allocators.

Overall, \toolname{} outperforms all baselines in terms of confirmed bugs while maintaining competitive precision.

\subsection{Summary Quality}

To evaluate summary generation, we construct a human-labeled dataset of functions drawn from the subject projects, partitioned into three ground-truth categories: (1)~\emph{MM\,+\,Reachable}: custom MM functions whose allocation or deallocation is on a feasible control-flow path; (2)~\emph{MM\,+\,Not\,Reachable}: custom MM functions that contain allocation or deallocation calls, but the call site lies on an infeasible path (e.g., guarded by an unsatisfiable condition); and (3)~\emph{Not\,MM}: functions that are not memory-management functions. This three-way split allows us to evaluate the LLM and Z3 components separately: the LLM should classify both MM categories as MM functions and reject Not\,MM; Z3 should retain MM\,+\,Reachable summaries and discard MM\,+\,Not\,Reachable summaries.

\noindent
\textbf{Dataset Construction.}
The full candidate pool comprises 88{,}474 functions that pass the pointer-type pre-filter (Phase~2 in Stage~1) across the eight subject projects. Applying Cochran's finite-population sample-size formula~\cite{DBLP:books/wi/Cochran77} at 95\% confidence and 5\% margin of error yields $n_{\mathrm{adj}} = 383$ samples per comparison side (full derivation in Appendix~\ref{appendix:sampling}). Because our evaluation involves two comparisons, namely (i)~MM vs.\ Not\,MM for evaluating the LLM, and (ii)~MM\,+\,Reachable vs.\ MM\,+\,Not\,Reachable for evaluating Z3, we sample 383 functions from each of the three classes, with Not\,MM doubled ($383 \times 2 = 766$) to balance against the combined MM pool. Samples are allocated equally across the eight projects, with approximately 48 functions per project per class. This yields 383 MM\,+\,Reachable, 383 MM\,+\,Not\,Reachable, and 766 Not\,MM functions, for 1{,}532 functions in total. To mitigate subjective bias, each function was independently labeled by two authors with Cohen's $\kappa$ of 0.94 (``almost perfect''~\cite{viera2005understanding}); disagreements were resolved through discussion.

\input{tab_summary_precision}

\noindent
\textbf{Results and Analysis.}
We evaluate using precision and recall, and compare against LeakGuard~\cite{liang2025leakguard}, a state-of-the-art approach that infers custom MM functions by mining allocation--deallocation usage patterns from the codebase.
We also run LeakGuard on the same eight repositories to obtain its inferred MM functions, and evaluate its precision and recall on the same set of 1{,}532 sampled functions.

Table~\ref{tab:summary_precision} reports precision and recall under two evaluation criteria.
\textit{MM} evaluates whether each approach correctly classifies functions as memory-management vs.\ non-MM: a true positive is a ground-truth MM function (MM\,+\,Reachable or MM\,+\,Not\,Reachable) correctly identified as MM; a false positive is a Not\,MM function mislabeled as MM; a false negative is an MM function the approach fails to identify.
\textit{Validated MM} evaluates a stricter criterion: whether each approach correctly identifies MM functions whose allocation or deallocation is on a feasible path.
For LeakGuard, both criteria are evaluated directly on its output, as it relies solely on NLP-based classification without symbolic path validation.
For \toolname{}, the MM criterion evaluates the LLM component alone (Phase~2), while the Validated MM criterion evaluates the full LLM\,+\,Z3 pipeline (Phases~2--3).

The results confirm that \toolname{} achieves 92.0\% MM precision with 100\% recall, significantly outperforming LeakGuard (78.1\% precision, 59.9\% recall).
After Z3 validation, \toolname{} achieves 100\% precision and recall on Validated MM, demonstrating that Z3 effectively filters the LLM's false positives while preserving all true MM functions with feasible allocation paths.

\subsection{Component Contribution}
\input{tab_reduction_summaries}
\input{tab_reduction_bugs}
Tables~\ref{tab:reduction_summaries} and~\ref{tab:reduction_bugs} quantify how each component contributes to the overall pipeline.

\noindent \textbf{Stage~1: Summary Generation (Table~\ref{tab:reduction_summaries}).} The pipeline progressively narrows from 99{,}648 extracted functions to 5{,}382 validated summaries of MM functions, achieving a 94.6\% overall reduction. Pre-filtering removes 11.2\% of functions that lack pointer types in their signatures. The LLM then classifies the remaining 88{,}474 candidates, rejecting 82.8\% as non-MM functions and producing 15{,}198 summaries. Z3 validation further eliminates 64.6\% of these summaries that lack feasible allocation or deallocation paths, retaining only validated summaries.

 \noindent \textbf{Stages~2--3: Bug Detection (Table~\ref{tab:reduction_bugs}).} 
Static analysis with injected summaries produces over 4{,}500 warnings per analyzer. 
Z3 feasibility filtering eliminates 86.5\% (CodeQL) and 83.7\% (Infer) of these warnings by discarding those on infeasible paths or that cannot be encoded.
LLM-based validation further reduces the remaining warnings by approximately 90--91\%, retaining 87 (CodeQL) and 67 (Infer) validated warnings.
\rw{Among these, manual validation confirms 44 true bugs for CodeQL and 31 for Infer, corresponding to precisions of 50.6\% and 46.3\%, respectively.}

In summary, the false positive reduction rate exceeds 98\% for both analyzers, with Z3 filtering removing most false alarms before LLM invocation, significantly reducing unnecessary cost. To assess whether the LLM incorrectly filters out true bugs, we randomly sampled 50 warnings classified as non-bugs by the LLM; none were true bugs upon manual inspection, suggesting that LLM filtering is unlikely to discard genuine bugs, consistent with prior work~\cite{du2026reducing}.

\subsection{Efficiency}

Table~\ref{tab:cost_time} presents the cost and runtime breakdown of \toolname{}. Summary generation dominates both cost and runtime (\$87.01 in total, approximately \$10.9 per program, and about 2.1 hours per program), as it must process all candidate functions across the codebase. \rw{Warning validation for each analyzer’s output incurs minimal overhead (\$4.50 for CodeQL and \$3.42 for Infer), due to the aggressive filtering performed by Z3 in earlier stages.} A key advantage of this design is that summary generation is a one-time cost; once generated and validated, summaries can be reused across multiple analyzers, project versions, and repeated analyses without regeneration. \rw{In practice, summaries only need to be regenerated when significant code changes occur (e.g., new custom allocators or major API refactoring); incremental updates between releases require no re-generation.} The static analysis phases (Phases 1, 3, 4, and 5) require approximately 32 minutes per program for CodeQL and 37 minutes for Infer on average, confirming that the pipeline's recurring cost is modest. \rw{Compared to LeakGuard, which averages 182 minutes per program, \toolname{} achieves a 4.9--5.7$\times$ speedup in the static analysis phase.}

\input{tab_cost_time}

\vspace{-0.1cm}
\section{Related Work}



\noindent\textbf{Memory Leak Bug Detection}
Static and dynamic techniques have been extensively studied for memory leak detection. Static approaches include abstract interpretation and model checking (e.g., Sparrow~\cite{jung2008practical}, Infer~\cite{fb_infer_github}), Boolean satisfiability modeling (SATURN~\cite{xie2005context}), and symbolic execution that tracks object lifecycles along program paths (KLEE~\cite{cadar2008klee}). A prominent line of work reduces leak detection to graph reachability on sparse value-flow graphs (SVFGs), as in Saber~\cite{DBLP:journals/tse/SuiYX14}, FastCheck~\cite{cherem2007practical}, Pinpoint~\cite{shi2018pinpoint}, LeakFix~\cite{gao2015safe}, and SMOKE~\cite{8812075}. Other static frameworks target specific defect patterns: partial call-path analysis (PCA~\cite{li2020pca}), intraprocedural inconsistencies in error-handling code (Hector~\cite{saha2013hector}), and practical memory ownership models (Clouseau~\cite{heine2003practical}). On the dynamic side, runtime monitors such as Purify~\cite{hastings1992purify}, AddressSanitizer~\cite{AddressSanitizer}, and Valgrind's memcheck~\cite{nethercote2007valgrind} instrument code to track memory states during execution or fuzzing~\cite{11391877,10.1145/3551349.3561161,MemLock}, while Sniper~\cite{jung2014automated} leverages hardware-assisted instruction sampling for lower overhead.
A separate line of work targets project-specific custom memory management (MM) functions through heuristics, NLP, or usage-aware rules (e.g., Goshawk~\cite{lyu2022goshawk}, K-MELD~\cite{299890}, LeakGuard~\cite{liang2025leakguard}, and others~\cite{242050,6671326,liu2014pf,liu2015pairminer}). However, these rely on surface-level signals (e.g., naming conventions, comments) and do not verify whether the inferred allocation or deallocation operations are reachable under the function's control flow. CMASan~\cite{hong2025cmasan} instead makes AddressSanitizer aware of custom allocators by instrumenting them dynamically, which requires inputs that exercise the code. In contrast, \toolname{} works statically, without test inputs, and combines LLM-based annotation with Z3-based symbolic validation to verify the reachability of inferred annotations and filter out infeasible leak paths.

\noindent
\noindent\textbf{LLM-assisted Static Analysis}
Large Language Models (LLMs) have been integrated into static analysis to bridge the semantic gap between code structures and developer intent. One line of work focuses on automated specification and annotation inference: frameworks such as IRIS~\cite{li2024iris} and LAMeD~\cite{shemetova2025lamed} extract source and sink metadata to guide external analyzers and mitigate path explosion, while SpecGen~\cite{ma2025specgen} and AutoSpec~\cite{wen2024enchanting} leverage LLMs to synthesize more general program specifications. A second line targets the high false discovery rates of classical tools through contextual warning triage, as exemplified by post-refinement systems like BUGLENS~\cite{li2025towards} and ZeroFalse~\cite{iranmanesh2025zerofalse}. More recently, the frontier has shifted toward autonomous agentic auditing, where agents such as RepoAudit~\cite{guo2025repoaudit} and QRS~\cite{tsigkourakos2026qrs} mimic human code auditors by navigating repositories on-demand to track data-flow facts across feasible paths.
Most closely related, IMMI~\cite{liu2024} detects kernel memory bugs by using an LLM to infer inconsistent caller/callee MM intentions on error-handling paths within a bespoke LLVM-IR engine. \toolname{} differs in three respects: (i) it produces analyzer-agnostic allocator/deallocator summaries that are Z3-validated for reachability and injected into off-the-shelf analyzers (CodeQL and Infer) for user-space C/C++; (ii) it adds a dedicated Z3 path-feasibility filter on the analyzer's output, which IMMI does not perform; and (iii) it frames memory leak detection as a multi-stage neuro-symbolic pipeline in which symbolic validation gates every LLM-generated artifact before it influences the analysis.

\section{Limitations and Future Work} \label{sec:limitation}  \noindent\textbf{Abstraction of the Symbolic Layer.} \toolname{}'s Z3 encoding is deliberately lightweight: it models control-flow reachability and the propagation of three memory-state predicates (\textit{alloc}, \textit{freed}, \textit{escaped}), rather than the full heap semantics of weakest-precondition or symbolic-execution engines. Branch conditions are treated as uninterpreted Boolean variables (identical tests share one variable), so the solver discards a warning only when no control-flow path with consistent branch choices reaches the leak while leaving the object allocated, unfreed, and unescaped; it does not refute paths that are infeasible solely under the numeric or pointer semantics of their guards, and, as the CFG is acyclic, it does not reason about loop-carried leaks. This keeps the per-warning cost negligible and suffices for the error-handling and wrapper-mediated leaks that dominate our study (Figures~\ref{fig:motivating-example-openssl}–\ref{fig:motivating-example-freerdp}), making Phase~5 an over-approximation; residual false positives that depend on richer semantics are deferred to the LLM validation in Phase~6.

\noindent\textbf{Scope and Dependencies.} \toolname{} currently targets memory leaks in C/C++ source parseable by Tree-sitter and buildable by CodeQL or Infer; related defects such as use-after-free and double-free are left to future work. Within this scope, Z3 validation discards structurally invalid summaries and infeasible warnings, but cannot detect a semantically incorrect summary that nonetheless admits a feasible path; such cases rely on the final LLM validation. Detection also depends on the specific LLM and prompts, affecting reproducibility when the model is not openly available. We mitigate this with Z3 validation of every summary and a second LLM pass on surviving warnings, but the dependence remains.
 The complete set of real leaks in our subjects is unknown, so we cannot report a false-negative rate. Leaks can be missed where the underlying analyzer reports no warning despite correct summaries, and where Phase~5 discards a warning it cannot encode (Section~\ref{subsec:stage3}).

\noindent\textbf{Future Work.} First, we plan to strengthen the symbolic layer with bounded loop unrolling, interpreted branch conditions to refute semantically infeasible paths, and a partitioned memory model for aliasing and heap updates. Second, since the allocator/deallocator summary abstraction is not specific to leaks, we plan to extend it to use-after-free, double-free, and other resource types (e.g., file descriptors, locks), as well as to other languages. Third, building on the validated leak paths, we plan to explore LLM-based program repair~\cite{le2019automated,fan2023automated,zhang2024autocoderover,10623236} for candidate fixes. Finally, we will study the impact of LLM choice, including smaller open-weight models, on cost and reproducibility.

\section{Conclusion}

This paper presents \toolname{}, a neuro-symbolic pipeline for detecting memory leaks in C/C++ programs. \toolname{} uses an LLM to infer the memory-management roles of custom functions, producing function summaries that extend the analyzer's built-in knowledge beyond standard primitives. Each summary is validated by Z3 to ensure the claimed allocation or deallocation is reachable on a feasible path. The validated summaries are injected into CodeQL and Infer, and Z3-based path feasibility filtering combined with LLM-based validation suppresses false positives. Evaluated on eight widely used open-source projects totaling over 3.6M lines of code, \toolname{} discovers 54 unique memory leaks (all confirmed or fixed), substantially outperforming vanilla CodeQL (19), vanilla Infer (3), Semgrep (2), and LeakGuard (20). 

\vspace{-0.1cm}
\section*{Ethical Considerations}
\label{sec:ethics}
This work reports previously unknown memory leaks in open-source software. All 54 bugs were disclosed to the upstream maintainers through each project's reporting channel before publication, and all have been confirmed or fixed. Memory leaks mainly cause resource exhaustion rather than code execution, which limits the potential for misuse. The study analyzes only public source code and involves no human subjects or personal data; therefore, no institutional ethics review was required.

\vspace{-0.1cm}
\section*{Acknowledgments}
\label{sec:ack}
This research/project is supported by the National Research Foundation, Singapore under its Artificial Intelligence (AI)-for-Science (AI4S) Challenge (Award ID: NRF-AI4SCH-2025-0003).
Any opinions, findings and conclusions or recommendations expressed in this material are those of the author(s) and do not reflect the views of National Research Foundation, Singapore.

\bibliographystyle{IEEEtran}
\bibliography{reference}

\appendices
\section{Prompt Templates}
\label{appendix:prompts}

This appendix lists the Phase~2 LLM summary generation prompt and the Phase~6 LLM warning validation prompt. To reduce latency, we follow the batching strategy of Li et al.~\cite{li2024iris}: Phase~2 batches 20 functions into a single prompt, while Phase~6 validates one function per call. Further details on the implementation of \toolname{}, together with instructions for running it, are available at: \url{https://github.com/jiekeshi/MemHint}.

\subsection{Phase~2: LLM Summary Generation Prompt}
\label{appendix:prompt-classify}

\begin{lstlisting}[basicstyle=\ttfamily\footnotesize,breaklines=true,frame=single,columns=fullflexible]
You are a memory safety expert analyzing C/C++ code to identify function semantics that help static analyzers detect memory leaks.

## Task
You will be given **one or more C/C++ functions** in a single request.
For **each function independently**, determine whether it is a memory allocator, deallocator, or neither.

The input will contain a block for each function, in this form:

```text
### Function: <func_name>
Return type: <return_type>
Parameters: <parameters>

```c
<code>
```

<optional context and Z3 feedback>
```

All function blocks will be concatenated into `{functions_block}`.

## Semantic Categories

### Allocator
Function returns **newly allocated heap memory** that caller must eventually free.

**Positive indicators:**
- Calls malloc/calloc/realloc/aligned_alloc/new/new[] and returns the result
- Calls another known allocator (e.g., g_malloc, xmalloc, kmalloc) and returns result
- Returns result of a wrapper function that allocates

**Negative indicators (NOT an allocator):**
- Returns pointer to static/global buffer
- Returns pointer to struct field or array member
- Returns one of the input arguments
- Allocates internally but doesn't return the allocated memory
- Returns stack-allocated memory (dangling pointer bug, but not allocator semantic)

### Deallocator
Function **frees/releases memory** passed as an argument.

**Positive indicators:**
- Calls free/delete/delete[]/g_free/kfree on an argument
- Calls another deallocator on an argument
- Wrapper around resource cleanup

## Analysis Guidelines

1. **Trace data flow:** Follow where return values come from and where arguments flow to.
2. **Consider all paths:** Check all branches and return statements.
3. **Indirect calls matter:** If function calls helper that allocates/frees, propagate that semantic.
4. **Be precise:** Only report semantics you can verify from the code.

## Output Format

Return a single JSON object whose **top-level keys are function names**.
For each function name, the value is an object with a `hints` array.

Each hint is a function summary with:
- `name`: the function name
- `role`: "Allocator" or "Deallocator"
- `target`: "return" for allocators (return value carries heap ownership), or "argN" for deallocators (the N-th argument is freed, 0-indexed)

### Example with two functions `foo` and `bar`

```json
{
  "foo": {
    "hints": [
      {"name": "foo", "role": "Allocator", "target": "return"},
      {"name": "foo", "role": "Deallocator", "target": "arg0"}
    ]
  },
  "bar": {
    "hints": []
  }
}
```

If no memory semantics apply to a function, still include that function name with `"hints": []`.

Now analyze **all functions in `{functions_block}`** and return the JSON result.
\end{lstlisting}

\subsection{Phase~6: LLM Warning Validation Prompt}
\label{appendix:prompt-validate}

\begin{lstlisting}[basicstyle=\ttfamily\footnotesize,breaklines=true,frame=single,columns=fullflexible]
You are a memory-safety expert. Analyze the following C function and the reported bug(s).

**Project:** {project_name}
**File:** {file}
**Function:** {function}
**Reported category:** {category}

**Reported issues (numbered 1, 2, ... for reference):**
  {idx}. Line {line_num}: {msg}
    allocation_site: {allocation_site}
    trace step {i}: {location}
    code at that step:
      {code}
    code at line {line_num}:
      {code_at_line}

**Function source:**
```c
{source with "// <-- reported bug" markers on bug lines}
```

Role: You are a senior static-analysis engineer specializing in C/C++ memory-safety.
You review memory-leak bug reports and assess whether the reported findings correspond to actual memory-leak defects in the program.

Does this function actually have a {bug_type_desc}? Determine whether the reported issue is a genuine bug (true) or a false alarm (false).

Decision policy:
  - true: at least one reported issue plausibly corresponds to a real memory leak based on the shown code.
  - false: all reported issues are not real defects based on the shown code.

If only some issues are real, output true and list only the real ones by index (1 = first, 2 = second, ...).

Respond with a single JSON object, no other text. Keep reason SHORT:
  {"verdict": true | false, "confidence": 0.0-1.0, "reason": "one short sentence", "bug_indices": [1] or [2,3] or []}

Output rules:
  - bug_indices: 1-based indices of reported issues you consider real; [] when verdict=false.
  - reason: ONE short sentence only. Do not quote code.
\end{lstlisting}

\section{Example Summary and Analyzer Injection}
\label{appendix:summary-example}

We illustrate the output of Stage~1 and the format in which it is injected into the two analyzers, using the FreeRDP running example from Section~II. After LLM classification and Z3 validation, the pipeline emits the following entries in its consolidated \texttt{\small hints.json}:

\begin{lstlisting}[basicstyle=\ttfamily\footnotesize,breaklines=true,frame=single,columns=fullflexible]
{
  "hints": {
    "freerdp_certificate_clone": [
      {
        "name": "freerdp_certificate_clone",
        "role": "Allocator",
        "target": "return"
      }
    ],
    "freerdp_certificate_new": [
      {
        "name": "freerdp_certificate_new",
        "role": "Allocator",
        "target": "return"
      }
    ],
    "freerdp_certificate_free": [
      {
        "name": "freerdp_certificate_free",
        "role": "Deallocator",
        "target": "arg0"
      }
    ]
  }
}
\end{lstlisting}

\noindent\textbf{CodeQL injection (data-extension YAML).} Each validated summary is appended to a data-extension pack that registers the function with CodeQL's built-in \texttt{\small allocationFunctionModel} or \texttt{\small deallocationFunctionModel}:

\begin{lstlisting}[basicstyle=\ttfamily\footnotesize,breaklines=true,frame=single,columns=fullflexible]
extensions:
  - addsTo:
      pack: codeql/cpp-all
      extensible: allocationFunctionModel
    data:
      - ["", "", false, "freerdp_certificate_clone", "", "", "", true]
      - ["", "", false, "freerdp_certificate_new", "", "", "", true]
  - addsTo:
      pack: codeql/cpp-all
      extensible: deallocationFunctionModel
    data:
      - ["", "", false, "freerdp_certificate_free", "0"]
\end{lstlisting}

\noindent Each row is a positional tuple for CodeQL's extensible predicates
\texttt{\small allocationFunctionModel(namespace, type, subtypes, name, sizeArg, sizeMult, reallocArg, requiresDealloc)}
and
\texttt{\small deallocationFunctionModel(namespace, type, subtypes, name, freedArg)}.
For these FreeRDP global functions, \texttt{\small namespace} and
\texttt{\small type} are empty, and \texttt{\small subtypes} is
\texttt{\small false}. The size and realloc slots are left empty because
our summaries encode ownership transfer but not allocation-size semantics.
The trailing \texttt{\small true} marks the returned allocation as requiring
a matching deallocation; the deallocator's trailing \texttt{\small "0"}
indicates that the first argument is freed, corresponding to
\texttt{\small arg0} in the summary.

\noindent\textbf{Infer injection (Pulse CLI flags).} For Infer, validated summaries are passed to the Pulse engine via two regex patterns at invocation time:

\begin{lstlisting}[basicstyle=\ttfamily\footnotesize,breaklines=true,frame=single,columns=fullflexible]
infer run --keep-going --pulse-only \
  --pulse-model-alloc-pattern \
    '^\(freerdp_certificate_clone\|freerdp_certificate_new\|...\)$' \
  --pulse-model-free-pattern \
    '^\(freerdp_certificate_free\|...\)$' \
  -- <build command>
\end{lstlisting}

\noindent Pulse then treats calls matching the allocation pattern as returning
caller-owned allocated objects, and calls matching the free pattern as releasing ownership of their argument. In this way, Infer can track memory ownership through project-specific wrappers using the validated summaries.

\section{Sample-Size Derivation}
\label{appendix:sampling}

For a finite population of size $N$, the required sample size at confidence level $Z$, worst-case proportion $p$, and margin of error $e$ is
\[
n = \frac{Z^{2}\,p(1-p)}{e^{2}}, \qquad
n_{\mathrm{adj}} = \frac{n}{1 + (n-1)/N},
\]
where the second equation applies the finite-population correction. Substituting $Z = 1.96$ (95\% confidence), $p = 0.5$ (worst-case proportion), $e = 0.05$ (5\% margin of error), and $N = 88{,}474$ (the candidate pool after pointer-type pre-filtering across the eight subject projects) gives an initial sample size of $n = 384.16$ and a corrected sample size of $n_{\mathrm{adj}} = 382.50$, which rounds to 383 samples per comparison side.

\end{document}

%% file: z3_validation_figure.tex
\begin{figure}[t!]
\centering

\begin{minipage}[t]{0.48\linewidth}
\centering
\textbf{(a) Valid Allocator}

\vspace{4pt}
\begin{tikzpicture}[
  node distance=0.4cm and 0.5cm,
  every node/.style={font=\scriptsize},
  block/.style={rectangle, draw, rounded corners=2pt, minimum width=0.9cm, minimum height=0.35cm, inner sep=2pt},
  branch/.style={diamond, draw, aspect=2.2, inner sep=1pt, fill=cyan!20},
  alloc/.style={block, fill=green!20},
  retok/.style={block, fill=yellow!30},
  ret/.style={block, fill=orange!20},
  entry/.style={block, fill=gray!15},
  edge/.style={-{Stealth[length=3pt]}, thick},
  tlabel/.style={font=\tiny, fill=white, inner sep=1pt},
]
  \node[entry] (e) {Entry};
  \node[branch, below=of e] (b) {$b_1$};
  \node[alloc, below left=0.4cm and 0.3cm of b] (a) {Alloc};
  \node[ret, below right=0.4cm and 0.3cm of b] (rn) {Ret \textsc{null}};
  \node[retok, below=of a] (rp) {Ret \texttt{ptr}};

  \draw[edge] (e) -- (b);
  \draw[edge] (b) -- node[tlabel, left]{\tiny T} (a);
  \draw[edge] (b) -- node[tlabel, right]{\tiny F} (rn);
  \draw[edge] (a) -- (rp);

  \draw[edge, blue!70, line width=1.2pt, dashed]
    (e.south) -- (b.north);
  \draw[edge, blue!70, line width=1.2pt, dashed]
    (b.south west) -- (a.north);
  \draw[edge, blue!70, line width=1.2pt, dashed]
    (a.south) -- (rp.north);

  \node[below=0.15cm of rp, font=\scriptsize\bfseries, blue!70] {Z3: \textsc{Sat} \checkmark};
\end{tikzpicture}
\end{minipage}
\hfill
\begin{minipage}[t]{0.48\linewidth}
\centering
\textbf{(b) Rejected Allocator}

\vspace{4pt}
\begin{tikzpicture}[
  node distance=0.4cm and 0.5cm,
  every node/.style={font=\scriptsize},
  block/.style={rectangle, draw, rounded corners=2pt, minimum width=0.9cm, minimum height=0.35cm, inner sep=2pt},
  branch/.style={diamond, draw, aspect=2.2, inner sep=1pt, fill=cyan!20},
  alloc/.style={block, fill=green!20},
  freenode/.style={block, fill=red!20},
  ret/.style={block, fill=orange!20},
  entry/.style={block, fill=gray!15},
  edge/.style={-{Stealth[length=3pt]}, thick},
  tlabel/.style={font=\tiny, fill=white, inner sep=1pt},
]
  \node[entry] (e) {Entry};
  \node[branch, below=of e] (b) {$b_1$};
  \node[alloc, below left=0.4cm and 0.3cm of b] (a) {Alloc};
  \node[ret, below right=0.4cm and 0.3cm of b] (rn) {Ret \textsc{null}};
  \node[freenode, below=of a] (f) {Free};
  \node[ret, below=of f] (rn2) {Ret \textsc{null}};

  \draw[edge] (e) -- (b);
  \draw[edge] (b) -- node[tlabel, left]{\tiny T} (a);
  \draw[edge] (b) -- node[tlabel, right]{\tiny F} (rn);
  \draw[edge] (a) -- (f);
  \draw[edge] (f) -- (rn2);

  \draw[edge, red!70, line width=1.2pt, dashed]
    (e.south) -- (b.north);
  \draw[edge, red!70, line width=1.2pt, dashed]
    (b.south west) -- (a.north);
  \draw[edge, red!70, line width=1.2pt, dashed]
    (a.south) -- (f.north);
  \draw[edge, red!70, line width=1.2pt, dashed]
    (f.south) -- (rn2.north);

  \node[below=0.15cm of rn2, font=\scriptsize\bfseries, red!70] {Z3: \textsc{Unsat} \ding{55}};
\end{tikzpicture}
\end{minipage}

\vspace{10pt}

\begin{minipage}[t]{0.48\linewidth}
\centering
\textbf{(c) Valid Deallocator}

\vspace{4pt}
\begin{tikzpicture}[
  node distance=0.4cm and 0.5cm,
  every node/.style={font=\scriptsize},
  block/.style={rectangle, draw, rounded corners=2pt, minimum width=0.9cm, minimum height=0.35cm, inner sep=2pt},
  branch/.style={diamond, draw, aspect=2.2, inner sep=1pt, fill=cyan!20},
  freenode/.style={block, fill=red!20},
  ret/.style={block, fill=orange!20},
  entry/.style={block, fill=gray!15},
  edge/.style={-{Stealth[length=3pt]}, thick},
  tlabel/.style={font=\tiny, fill=white, inner sep=1pt},
]
  \node[entry] (e) {Entry};
  \node[branch, below=of e] (b) {$b_1$};
  \node[freenode, below left=0.4cm and 0.15cm of b] (f) {\texttt{free(arg0)}};
  \node[ret, below right=0.4cm and 0.15cm of b] (rn) {Return};
  \node[ret, below=of f] (re) {Return};

  \draw[edge] (e) -- (b);
  \draw[edge] (b) -- node[tlabel, left]{\tiny T} (f);
  \draw[edge] (b) -- node[tlabel, right]{\tiny F} (rn);
  \draw[edge] (f) -- (re);

  \draw[edge, blue!70, line width=1.2pt, dashed]
    (e.south) -- (b.north);
  \draw[edge, blue!70, line width=1.2pt, dashed]
    (b.south west) -- (f.north);
  \draw[edge, blue!70, line width=1.2pt, dashed]
    (f.south) -- (re.north);

  \node[below=0.15cm of re, font=\scriptsize\bfseries, blue!70] {Z3: \textsc{Sat} \checkmark};
\end{tikzpicture}
\end{minipage}
\hfill
\begin{minipage}[t]{0.48\linewidth}
\centering
\textbf{(d) Rejected Deallocator}

\vspace{4pt}
\begin{tikzpicture}[
  node distance=0.4cm and 0.5cm,
  every node/.style={font=\scriptsize},
  block/.style={rectangle, draw, rounded corners=2pt, minimum width=0.9cm, minimum height=0.35cm, inner sep=2pt},
  branch/.style={diamond, draw, aspect=2.2, inner sep=1pt, fill=cyan!20},
  freenode/.style={block, fill=red!20},
  ret/.style={block, fill=orange!20},
  entry/.style={block, fill=gray!15},
  edge/.style={-{Stealth[length=3pt]}, thick},
  tlabel/.style={font=\tiny, fill=white, inner sep=1pt},
]
  \node[entry] (e) {Entry};
  \node[branch, below=of e] (b) {$b_1$};
  \node[freenode, below left=0.4cm and 0.05cm of b] (f) {\texttt{free(arg0->f)}};
  \node[ret, below right=0.4cm and 0.15cm of b] (rn) {Return};
  \node[ret, below=of f] (re) {Return};

  \draw[edge] (e) -- (b);
  \draw[edge] (b) -- node[tlabel, left]{\tiny T} (f);
  \draw[edge] (b) -- node[tlabel, right]{\tiny F} (rn);
  \draw[edge] (f) -- (re);

  \draw[edge, red!70, line width=1.2pt, dashed]
    (e.south) -- (b.north);
  \draw[edge, red!70, line width=1.2pt, dashed]
    (b.south west) -- (f.north);
  \draw[edge, red!70, line width=1.2pt, dashed]
    (f.south) -- (re.north);

  \node[below=0.15cm of re, font=\scriptsize\bfseries, red!70, align=center] {%
    \texttt{arg0->f} $\neq$ alias of \texttt{arg0}\\
    Z3: \textsc{Unsat} \ding{55}};
\end{tikzpicture}
\end{minipage}

\caption{Z3-based Summary Validation.
(a)--(b) Allocator: (a) SAT: memory allocated at return; (b) UNSAT: memory is freed on all paths. 
(c)--(d) Deallocator: (c) SAT: target argument is freed; (d) UNSAT: only internal fields (e.g., \texttt{arg0->f}) are freed.}
\label{fig:z3-validation}
\end{figure}

%% file: z3_feasibility_figure.tex
\begin{figure}[t!]
\centering
\begin{adjustbox}{width=\columnwidth}

\begin{minipage}[t]{0.50\linewidth}
\centering
\textbf{Function Code (FreeRDP)}

\vspace{4pt}
% ...

\includegraphics[width=\linewidth]{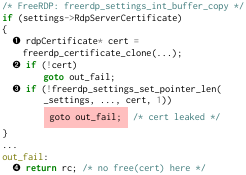}

\vspace{4pt}
{\tiny
\textbf{Key nodes:}
\ding{202}~Alloc $\to$
\ding{203}~Branch ($b_1$: \texttt{!cert}) $\to$
\ding{204}~Branch ($b_2$: \texttt{!set\_ptr}) $\to$
\ding{205}~Return
}
\end{minipage}%
\hspace{2pt}%
\begin{minipage}[t]{0.50\linewidth}
\centering
\textbf{CFG with Z3 State Encoding}

\vspace{4pt}
\begin{tikzpicture}[
  node distance=0.35cm and 0.4cm,
  every node/.style={font=\tiny},
  block/.style={rectangle, draw, rounded corners=2pt, minimum width=0.9cm, minimum height=0.3cm, inner sep=1.5pt},
  branch/.style={diamond, draw, aspect=2.2, inner sep=1pt, fill=cyan!20},
  alloc/.style={block, fill=green!20},
  exitnode/.style={block, fill=gray!30},
  entry/.style={block, fill=gray!15},
  edge/.style={-{Stealth[length=2.5pt]}, semithick},
  tlabel/.style={font=\tiny, fill=white, inner sep=0.5pt},
]
  \node[entry] (e) {Entry};
  \node[alloc, below=of e] (a) {\ding{202}\,Alloc: \texttt{cert}};
  \node[branch, below=of a] (b1) {\ding{203}\,$b_1$};
  \node[exitnode, right=0.7cm of b1] (ex1) {\ding{205}\,Ret};
  \node[branch, below=of b1] (b2) {\ding{204}\,$b_2$};
  \node[exitnode, right=0.7cm of b2] (ex2) {\ding{205}\,Ret};
  \node[exitnode, below=of b2] (ex3) {Ret};

  \draw[edge] (e) -- (a);
  \draw[edge] (a) -- (b1);
  \draw[edge] (b1) -- node[tlabel, above]{\tiny T} (ex1);
  \draw[edge] (b1) -- node[tlabel, left]{\tiny F} (b2);
  \draw[edge] (b2) -- node[tlabel, above]{\tiny T} (ex2);
  \draw[edge] (b2) -- node[tlabel, left]{\tiny F} (ex3);

  \draw[edge, red!70, line width=1.2pt, dashed]
    (e.south) -- (a.north);
  \draw[edge, red!70, line width=1.2pt, dashed]
    (a.south) -- (b1.north);
  \draw[edge, red!70, line width=1.2pt, dashed]
    (b1.south) -- (b2.north);
  \draw[edge, red!70, line width=1.2pt, dashed]
    (b2.east) -- (ex2.west);

  \node[above=0.01cm of ex2, font=\tiny\itshape, red!70] {feasible path};

  \node[below=0.25cm of ex3, font=\tiny\bfseries, red!70, align=center] {%
    $\neg b_1 \wedge b_2$: Alloc $\to$ Ret (no free)\\
    alloc=T, freed=F, escaped=F\\
    Z3: \textsc{Sat} --- \textbf{feasible leak}};

\end{tikzpicture}
\end{minipage}

\end{adjustbox}

\caption{
Z3-based memory-leak feasibility checking (Phase~5) on the FreeRDP bug in Figure~\ref{fig:motivating-example-freerdp}.
Left: flagged function with key CFG nodes (allocation \ding{202}, branches \ding{203}, \ding{204}, return \ding{205}).
Right: intra-procedural CFG with Z3 encoding.
Along the dashed path, \texttt{cert} is allocated ($\mathit{alloc}=\mathit{true}$) but neither freed nor escaped; Z3 returns \textsc{SAT}, indicating a feasible leak.
}
\label{fig:z3-feasibility}
\end{figure}

%% file: tab_results.tex
\begin{table*}[t]
\centering
\small
\renewcommand{\arraystretch}{1.1}
\begin{threeparttable}
\caption{
Bugs found by \toolname{} across programs.
}
\label{tab:results}
\begin{tabular}{l l l r r r r r r r}
\hline
& & & & &  \multicolumn{5}{c}{\textbf{\toolname{}}} \\
\cline{6-10}
\textbf{Program} &
\textbf{Type} &
\textbf{Version} &
\textbf{SLOC} &
\textbf{\#Func.} &
\textbf{CodeQL} &
\textbf{Infer} &
\textbf{Overlap} &
\textbf{Detected} &
\textbf{Confirmed} \\
\hline
Vim     & Text Editor        & 9.2.0015       & 429K  & 10.9K & 18 & 15 & 11 & 22 & 22  \\
Tmux    & Terminal Tool  & 3.6a           & 75K   & 2.5K  & 9  & 6  & 5  & 10 & 10  \\
OpenSSL & Crypto Library      & 3.6.1          & 725K  & 20.7K & 4  & 6  & 1  & 9  & 9   \\
Redis   & Database System      & 8.6-rc1        & 241K  & 5.2K  & 3  & 0  & 0  & 3  & 3   \\
FreeRDP & Remote Desktop    & 3.23.0         & 436K  & 13.2K & 5  & 2  & 2  & 5  & 5   \\
curl    & Network Tool & 8.19.0-rc3     & 201K  & 5.0K  & 1  & 1  & 1  & 1  & 1   \\
FFmpeg  & Multimedia Tool    & n8.1-dev       & 1.3M  & 31.1K & 2  & 1  & 1  & 2  & 2   \\
linux/staging & OS Subsystem & v7.0-rc1 & 229K & 6.3K & 2 & 0 & 0 & 2 & 2 \\
\hline
\textbf{Total} & -- & -- & \textbf{3.6M} & \textbf{94.9K} & \textbf{44} & \textbf{31} & \textbf{21} & \textbf{54} & \textbf{54}\\
\hline
\end{tabular}
\begin{tablenotes}[flushleft]
\scriptsize
\item \emph{SLOC}: source lines of code (excluding blanks/comments).
\emph{\#Func.}: number of functions.
\emph{CodeQL}/\emph{Infer}: bugs detected by each summary-augmented analyzer.
\emph{Overlap}: bugs found by both.
\emph{Detected}: union of validated bugs.
\emph{Confirmed}: bugs confirmed or fixed by maintainers.
\end{tablenotes}
\end{threeparttable}
\end{table*}

%% file: tab_baseline.tex
\begin{table}[t]
\centering
\small
\setlength{\tabcolsep}{2.4pt}
\renewcommand{\arraystretch}{1.05}
\caption{
Comparison of bugs detected per program between \toolname{} and baseline tools.
}
\label{tab:baseline_comparison}
\begin{tabular}{lrrrrr}
\hline
\textbf{Program} & \textbf{CodeQL$^\ast$} & \textbf{Infer} & \textbf{LeakGuard} & \textbf{Sgrep} & \textbf{\toolname{}} \\
\hline
Vim     & 10 & 3 & 9 & 0 & 22 \\
Tmux    & 7 & 0 & 4 & 1 & 10 \\
OpenSSL & 0 & 0 & 4 & 1 & 9 \\
Redis   & 0 & 0 & 2 & 0 & 3 \\
FreeRDP & 2 & 0 & 1 & 0 & 5 \\
curl    & 0 & 0 & 0 & 0 & 1 \\
FFmpeg  & 0 & 0 & 0 & 0 & 2 \\
linux/staging & 0 & 0 & 0 & 0 & 2 \\
\hline
\textbf{Total} & \textbf{19} & \textbf{3} & \textbf{20} & \textbf{2} & \textbf{54} \\
\hline
\multicolumn{6}{l}{\scriptsize{$^\ast$ same bugs with/without extended queries. Sgrep = Semgrep.}}
\end{tabular}
\end{table}

%% file: tab_summary_precision.tex
\begin{table}[t]
\centering
\small
\renewcommand{\arraystretch}{1.1}
\begin{threeparttable}
\caption{
Evaluation of MM identification and validation.
}
\label{tab:summary_precision}
\begin{tabular*}{\columnwidth}{@{\extracolsep{\fill}}l rr rr}
\hline
& \multicolumn{2}{c}{\textbf{MM}} & \multicolumn{2}{c}{\textbf{Validated MM}} \\
\cline{2-3} \cline{4-5}
& \textbf{Precision} & \textbf{Recall} & \textbf{Precision} & \textbf{Recall} \\
\hline
LeakGuard      & 78.1\% & 59.9\% & 34.0\% & 52.2\% \\
\toolname{}    & 92.0\% & 100.0\% & 100.0\% & 100.0\% \\
\hline
\end{tabular*}
\begin{tablenotes}[flushleft]
\scriptsize
\item \emph{MM}: MM vs.\ non-MM classification.
\emph{Validated MM}: reachable MM functions on feasible paths.
\end{tablenotes}
\end{threeparttable}
\end{table}

%% file: tab_reduction_summaries.tex
\begin{table}[t]
\centering
\small
\setlength{\tabcolsep}{4pt}
\renewcommand{\arraystretch}{1.05}
\begin{threeparttable}
\caption{
Contribution of each Stage~1 phase to candidate reduction and summary validation.
}
\label{tab:reduction_summaries}
\begin{tabular}{l r r r r}
\hline
\textbf{Program} &
\textbf{\#Extr.} &
\textbf{\#Cand.} &
\textbf{\#Summ.} &
\textbf{\#Valid.} \\
\hline
Vim     & 11,071 & 8,539 {\color{red}\tiny$\downarrow$22.9\%} & 2,539 {\color{red}\tiny$\downarrow$70.3\%} & 688 {\color{red}\tiny$\downarrow$72.9\%} \\
Tmux    & 2,612 & 2,528 {\color{red}\tiny$\downarrow$3.2\%} & 790 {\color{red}\tiny$\downarrow$68.8\%} & 467 {\color{red}\tiny$\downarrow$40.9\%} \\
OpenSSL & 24,068 & 20,996 {\color{red}\tiny$\downarrow$12.8\%} & 3,454 {\color{red}\tiny$\downarrow$83.5\%} & 1,740 {\color{red}\tiny$\downarrow$49.6\%} \\
Redis   & 5,660 & 4,882 {\color{red}\tiny$\downarrow$13.7\%} & 1,308 {\color{red}\tiny$\downarrow$73.2\%} & 493 {\color{red}\tiny$\downarrow$62.3\%} \\
FreeRDP & 13,940 & 12,160 {\color{red}\tiny$\downarrow$12.8\%} & 2,256 {\color{red}\tiny$\downarrow$81.4\%} & 1,161 {\color{red}\tiny$\downarrow$48.5\%} \\
curl    & 4,978 & 4,413 {\color{red}\tiny$\downarrow$11.3\%} & 736 {\color{red}\tiny$\downarrow$83.3\%} & 219 {\color{red}\tiny$\downarrow$70.2\%} \\
FFmpeg  & 29,705 & 27,930 {\color{red}\tiny$\downarrow$6.0\%} & 3,596 {\color{red}\tiny$\downarrow$87.1\%} & 494 {\color{red}\tiny$\downarrow$86.3\%} \\
linux/staging & 7,614 & 7,026 {\color{red}\tiny$\downarrow$7.7\%} & 519 {\color{red}\tiny$\downarrow$92.6\%} & 120 {\color{red}\tiny$\downarrow$76.9\%} \\
\hline
\textbf{Total} & \textbf{99,648} & \textbf{88,474} {\color{red}\tiny$\downarrow$11.2\%} & \textbf{15,198} {\color{red}\tiny$\downarrow$82.8\%} & \textbf{5,382} {\color{red}\tiny$\downarrow$64.6\%} \\
\hline
\end{tabular}
\begin{tablenotes}[flushleft]
\scriptsize
\item \emph{\#Extr.}: parsed functions/macros.
\emph{\#Cand.}: after pre-filtering.
\emph{\#Summ.}: LLM-generated summaries.
\emph{\#Valid.}: Z3-validated summaries.
{\color{red}Red} numbers show reduction per step.
\end{tablenotes}
\end{threeparttable}
\end{table}

%% file: tab_reduction_bugs.tex
\begin{table}[t]
\centering
\small
\setlength{\tabcolsep}{4pt}
\renewcommand{\arraystretch}{1.05}
\begin{threeparttable}
\caption{
Contributions of Stages~2--3 to warning filtering and validation.
}
\label{tab:reduction_bugs}
\begin{tabular}{l@{\hspace{3pt}}l@{\hspace{3pt}}r r r r}
\hline
\textbf{Program} &
&
\textbf{\#Warn.} &
\textbf{\#Z3} &
\textbf{\#Valid.} &
\textbf{\#Det.} \\
\hline
\multirow{2}{*}{Vim}
  & CodeQL & 1,011 & 86 {\color{red}\tiny$\downarrow$91.5\%} & 24 {\color{red}\tiny$\downarrow$72.1\%} & 18 {\color{blue}\tiny(75.0\%)} \\
  & Infer & 1,032 & 147 {\color{red}\tiny$\downarrow$85.8\%} & 25 {\color{red}\tiny$\downarrow$83.0\%} & 15 {\color{blue}\tiny(60.0\%)} \\
\hline
\multirow{2}{*}{Tmux}
  & CodeQL & 291 & 70 {\color{red}\tiny$\downarrow$75.9\%} & 13 {\color{red}\tiny$\downarrow$81.4\%} & 9 {\color{blue}\tiny(69.2\%)} \\
  & Infer & 304 & 36 {\color{red}\tiny$\downarrow$88.2\%} & 7 {\color{red}\tiny$\downarrow$80.6\%} & 6 {\color{blue}\tiny(85.7\%)} \\
\hline
\multirow{2}{*}{OpenSSL}
  & CodeQL & 2,547 & 289 {\color{red}\tiny$\downarrow$88.7\%} & 20 {\color{red}\tiny$\downarrow$93.1\%} & 4 {\color{blue}\tiny(20.0\%)} \\
  & Infer & 2,078 & 351 {\color{red}\tiny$\downarrow$83.1\%} & 27 {\color{red}\tiny$\downarrow$92.3\%} & 6 {\color{blue}\tiny(22.2\%)} \\
\hline
\multirow{2}{*}{Redis}
  & CodeQL & 705 & 89 {\color{red}\tiny$\downarrow$87.4\%} & 8 {\color{red}\tiny$\downarrow$91.0\%} & 3 {\color{blue}\tiny(37.5\%)} \\
  & Infer & 198 & 33 {\color{red}\tiny$\downarrow$83.3\%} & 0 {\color{red}\tiny$\downarrow$100\%} & 0 {\color{blue}\tiny(--)} \\
\hline
\multirow{2}{*}{FreeRDP}
  & CodeQL & 890 & 141 {\color{red}\tiny$\downarrow$84.2\%} & 6 {\color{red}\tiny$\downarrow$95.7\%} & 5 {\color{blue}\tiny(83.3\%)} \\
  & Infer & 541 & 82 {\color{red}\tiny$\downarrow$84.8\%} & 3 {\color{red}\tiny$\downarrow$96.3\%} & 2 {\color{blue}\tiny(66.7\%)} \\
\hline
\multirow{2}{*}{curl}
  & CodeQL & 403 & 47 {\color{red}\tiny$\downarrow$88.3\%} & 5 {\color{red}\tiny$\downarrow$89.4\%} & 1 {\color{blue}\tiny(20.0\%)} \\
  & Infer & 195 & 37 {\color{red}\tiny$\downarrow$81.0\%} & 1 {\color{red}\tiny$\downarrow$97.3\%} & 1 {\color{blue}\tiny(100\%)} \\
\hline
\multirow{2}{*}{FFmpeg}
  & CodeQL & 404 & 80 {\color{red}\tiny$\downarrow$80.2\%} & 2 {\color{red}\tiny$\downarrow$97.5\%} & 2 {\color{blue}\tiny(100\%)} \\
  & Infer & 184 & 42 {\color{red}\tiny$\downarrow$77.2\%} & 1 {\color{red}\tiny$\downarrow$97.6\%} & 1 {\color{blue}\tiny(100\%)} \\
\hline
\multirow{2}{*}{linux/staging}
  & CodeQL & 120 & 57 {\color{red}\tiny$\downarrow$52.5\%} & 9 {\color{red}\tiny$\downarrow$84.2\%} & 2 {\color{blue}\tiny(22.2\%)} \\
  & Infer & 30 & 14 {\color{red}\tiny$\downarrow$53.3\%} & 3 {\color{red}\tiny$\downarrow$78.6\%} & 0 {\color{blue}\tiny(0\%)} \\
\hline
\multirow{2}{*}{\textbf{Total}}
  & CodeQL & \textbf{6,371} & \textbf{859} {\color{red}\tiny$\downarrow$86.5\%} & \textbf{87} {\color{red}\tiny$\downarrow$89.9\%} & \textbf{44} {\color{blue}\tiny(50.6\%)} \\
  & Infer & \textbf{4,562} & \textbf{742} {\color{red}\tiny$\downarrow$83.7\%} & \textbf{67} {\color{red}\tiny$\downarrow$91.0\%} & \textbf{31} {\color{blue}\tiny(46.3\%)} \\
\hline
\end{tabular}
\begin{tablenotes}[flushleft]
\scriptsize
\item \emph{\#Warn.}: SAST tool output warnings.
\emph{\#Z3}: after Z3 filtering.
\emph{\#Valid.}: LLM-validated warnings.
\emph{\#Det.}: human-validated bugs ({\color{blue}precision}).
{\color{red}Red} percentages show reduction per step.
\end{tablenotes}
\end{threeparttable}
\end{table}

%% file: tab_cost_time.tex
\begin{table}[t]
\centering
\small
\renewcommand{\arraystretch}{1.1}
\begin{threeparttable}
\caption{
Cost and runtime breakdown of \toolname{} across programs.
}
\label{tab:cost_time}
\small
\begin{tabular*}{\columnwidth}{@{\extracolsep{\fill}}l l r r r r}
\hline
\textbf{Program} &
&
\textbf{Gen} &
\textbf{Vrf} &
\textbf{Static} &
\textbf{LeakGuard} \\
\hline
\multirow{2}{*}{Vim}
  & CodeQL & \multirow{2}{*}{10.20} & 0.46 & 56m & \multirow{2}{*}{1.4h} \\
  & Infer &                        & 0.65 & 57m & \\
\hline
\multirow{2}{*}{Tmux}
  & CodeQL & \multirow{2}{*}{2.27} & 0.29 & 3m & \multirow{2}{*}{31m} \\
  & Infer &                       & 0.17 & 3m & \\
\hline
\multirow{2}{*}{OpenSSL}
  & CodeQL & \multirow{2}{*}{17.00} & 1.55 & 1.3h & \multirow{2}{*}{7.1h} \\
  & Infer &                        & 1.52 & 1.9h & \\
\hline
\multirow{2}{*}{Redis}
  & CodeQL & \multirow{2}{*}{5.83} & 0.44 & 8m & \multirow{2}{*}{38m} \\
  & Infer &                       & 0.14 & 10m & \\
\hline
\multirow{2}{*}{FreeRDP}
  & CodeQL & \multirow{2}{*}{14.11} & 0.52 & 32m & \multirow{2}{*}{1.1h} \\
  & Infer &                        & 0.36 & 27m & \\
\hline
\multirow{2}{*}{curl}
  & CodeQL & \multirow{2}{*}{4.86} & 0.21 & 10m & \multirow{2}{*}{13m} \\
  & Infer &                       & 0.17 & 8m  & \\
\hline
\multirow{2}{*}{FFmpeg}
  & CodeQL & \multirow{2}{*}{26.09} & 0.30 & 48m & \multirow{2}{*}{10.2h} \\
  & Infer &                        & 0.17 & 1.1h & \\
\hline
\multirow{2}{*}{linux/staging}
  & CodeQL & \multirow{2}{*}{6.65} & 0.73 & 20m & \multirow{2}{*}{3.1h} \\
  & Infer &                       & 0.24 & 12m & \\
\hline
\multirow{2}{*}{\textbf{Total}}
  & CodeQL & \multirow{2}{*}{\textbf{87.01}} & \textbf{4.50} & \textbf{4.3h} & \multirow{2}{*}{\textbf{24.3h}} \\
  & Infer &                                 & \textbf{3.42} & \textbf{5.0h} & \\
\hline
\end{tabular*}
\begin{tablenotes}[flushleft]
\scriptsize
\item \emph{Gen}: LLM cost (\$) for summary generation.
\emph{Vrf}: LLM cost (\$) for warning validation.
\emph{Static}: \toolname{} static-phase runtime (Phases 1, 3, 4, and 5).
\end{tablenotes}
\end{threeparttable}
\end{table}